%% file: camera_ready.tex
\documentclass[9pt,sigconf,screen]{acmart}
\input{setting-acm}
\graphicspath{{./figs/}{../}}
\copyrightyear{2026}
\acmYear{2026}
\setcopyright{cc}
\setcctype{by}
\acmConference[MLCAD '26]{2026 ACM/IEEE International Symposium on Machine Learning for CAD}{September 07--09, 2026}{Jeju Island, Republic of Korea}
\acmBooktitle{2026 ACM/IEEE International Symposium on Machine Learning for CAD (MLCAD '26), September 07--09, 2026, Jeju Island, Republic of Korea}
\acmDOI{10.1145/3831599.3840341}
\acmISBN{979-8-4007-2878-5/2026/09}

\usepackage{graphbox}
\usepackage{enumitem}

\usepackage{mathtools}
\newcolumntype{P}[1]{>{\centering\arraybackslash}p{#1}}
\newcolumntype{M}[1]{>{\centering\arraybackslash}m{#1}}

\newcommand{\orfs}{OpenROAD-flow-scripts}

\newcommand{\hf}{H_f}

\newcommand{\zyz}[1]{\textcolor{black}{ #1}}
\usepackage{tikz}
\usepackage[pages=some]{background}

\begin{document}
\fontsize{8.8pt}{10.4pt}\selectfont

\title{From Tool Invocation to Source-Mechanism Exploration: Protected White-Box DSE for Open-Source EDA}

\author{Zhiyu Zheng}
\affiliation{%
  \institution{Fudan University}
  \city{Shanghai}
  \country{China}}
\email{zyzheng24@m.fudan.edu.cn}

\author{Yiming Du}
\affiliation{%
  \institution{Fudan University}
  \city{Shanghai}
  \country{China}}
\email{ymdu26@m.fudan.edu.cn}

\author{Ziyi Wang}
\affiliation{%
  \institution{The Chinese University of Hong Kong}
  \city{Hong Kong}
  \country{China}}
\email{zeayw1@gmail.com}

\author{Zhiang Wang}
\affiliation{%
  \institution{Fudan University}
  \city{Shanghai}
  \country{China}}
\email{zhiangwang@fudan.edu.cn}

\keywords{open-source EDA, white-box DSE, detailed placement, OpenROAD}

\begin{abstract}
Open-source EDA tools allow design-space exploration (DSE) to move beyond public knobs and into bounded source-level mechanisms inside staged optimizers. We present ReviewDSE, a protected white-box DSE framework that explores such mechanisms for a target design. ReviewDSE evaluates complete source candidates under a protected evaluator and records reusable search knowledge as reviewed mechanism-level evidence. It first constructs method evidence and source-start branches from calibration designs, then uses these fixed warm-start products to initialize target-case exploration under Teacher review and full-flow validation. We instantiate ReviewDSE on OpenROAD detailed placement as a representative staged open-source EDA optimizer. Across nine target tasks, ReviewDSE reduces final post-DPL half-perimeter wirelength (HPWL) by 1.78\% on average under a 2$\times$ runtime gate, compared with 0.38\% for public-knob black-box DSE. A runtime-aware ReviewDSE selection retains a 1.68\% reduction at 1.11$\times$ runtime, and full-flow review exposes stage-composability failures while source-mechanism exploration repairs hard cut-row legality failures.
\end{abstract}

\ccsdesc[500]{Hardware~Physical design (EDA)}
\ccsdesc[500]{Hardware~Placement}

\maketitle
\backgroundsetup{
  opacity=1,
  scale=1,
  angle=0,
  contents={
    \begin{tikzpicture}[remember picture, overlay]
      \node[
        anchor=north east,
        inner xsep=50pt,
        inner ysep=10pt
      ] at (current page.north east) {
        \href{https://www.acm.org/publications/policies/artifact-review-and-badging-current}{
          \includegraphics[width=50pt]{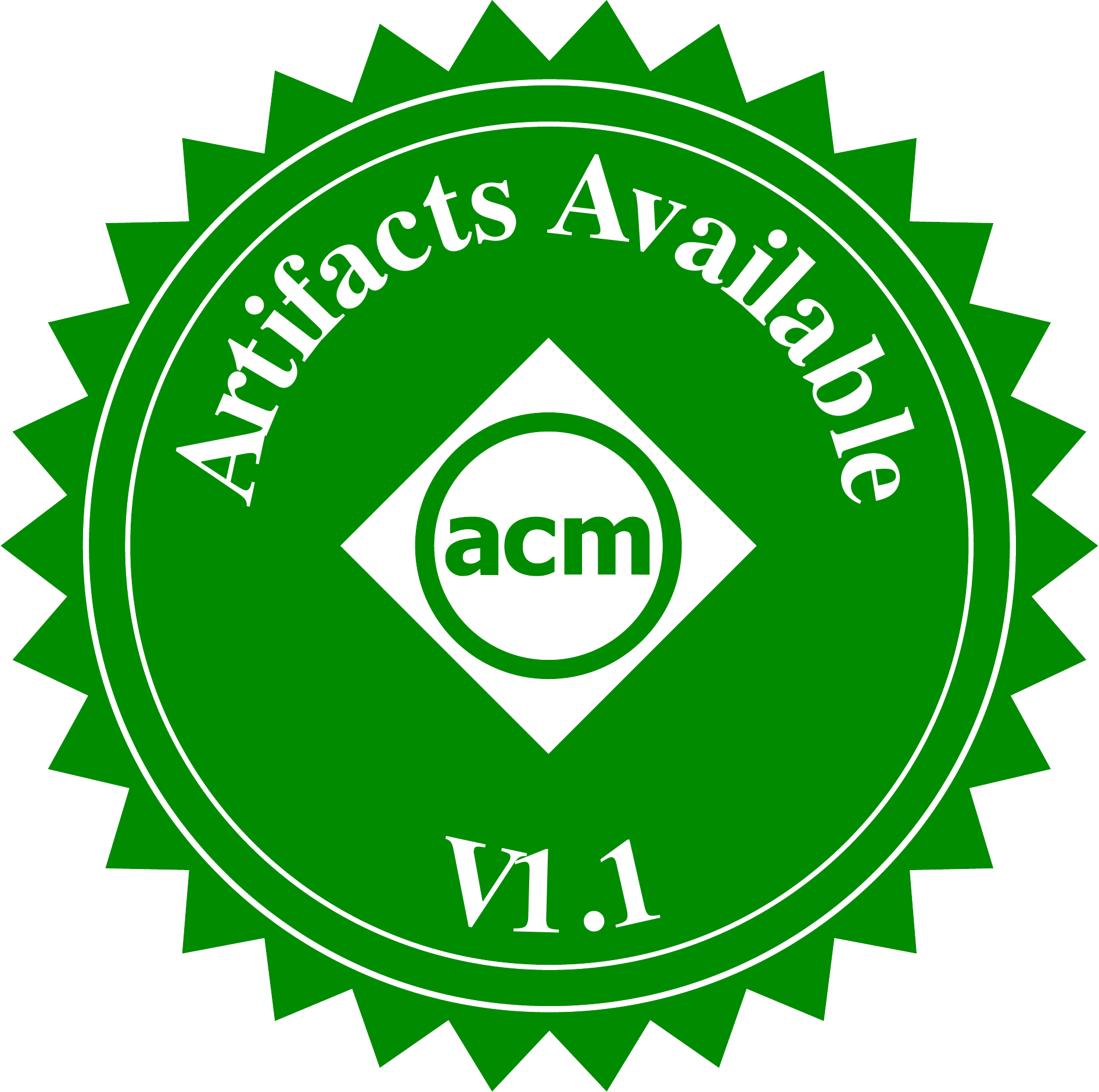}
          \includegraphics[width=50pt]{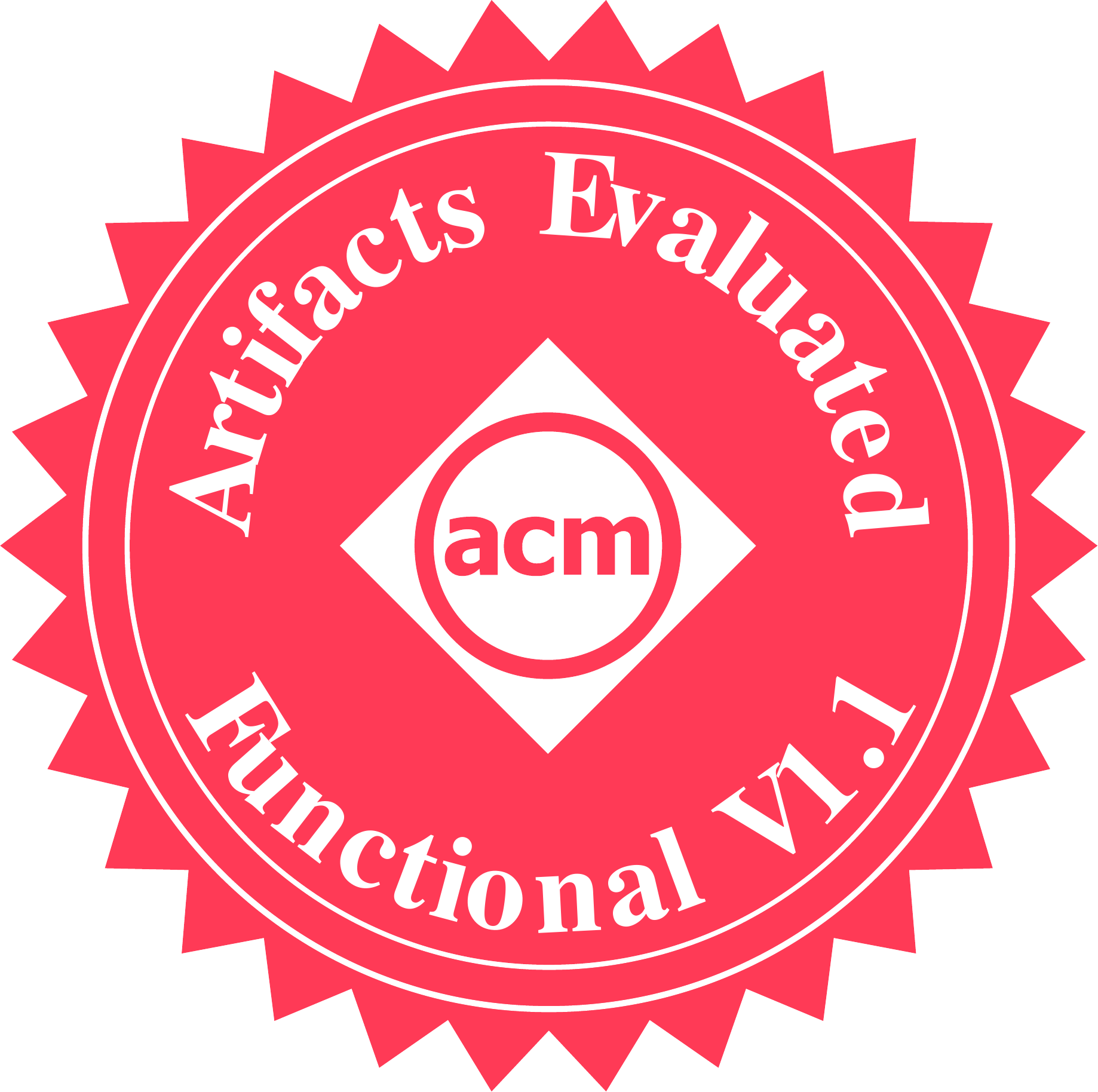}
        }
      };
    \end{tikzpicture}
  }
}
\BgThispage
\vspace{-0.6em}
\section{Introduction}

Design-space exploration (DSE) is a standard methodology for improving electronic-design-automation (EDA) flows~\cite{geng2022edaDSEsurvey}. Conventional DSE treats each tool as a fixed executable and searches its public control surface, such as command options, flow scripts, or other exposed knobs, to find configurations with better quality of results~\cite{ma2019cad}.
This black-box formulation is practical and broadly applicable, especially for closed-source tools, but it leaves quality-critical implementation choices inside large EDA codebases outside the search space.

\begin{figure}[tb]
    \subfloat[Conventional Black-box DSE]{
        \centering
        \label{subfig:white}
\includegraphics[width=.23\textwidth]{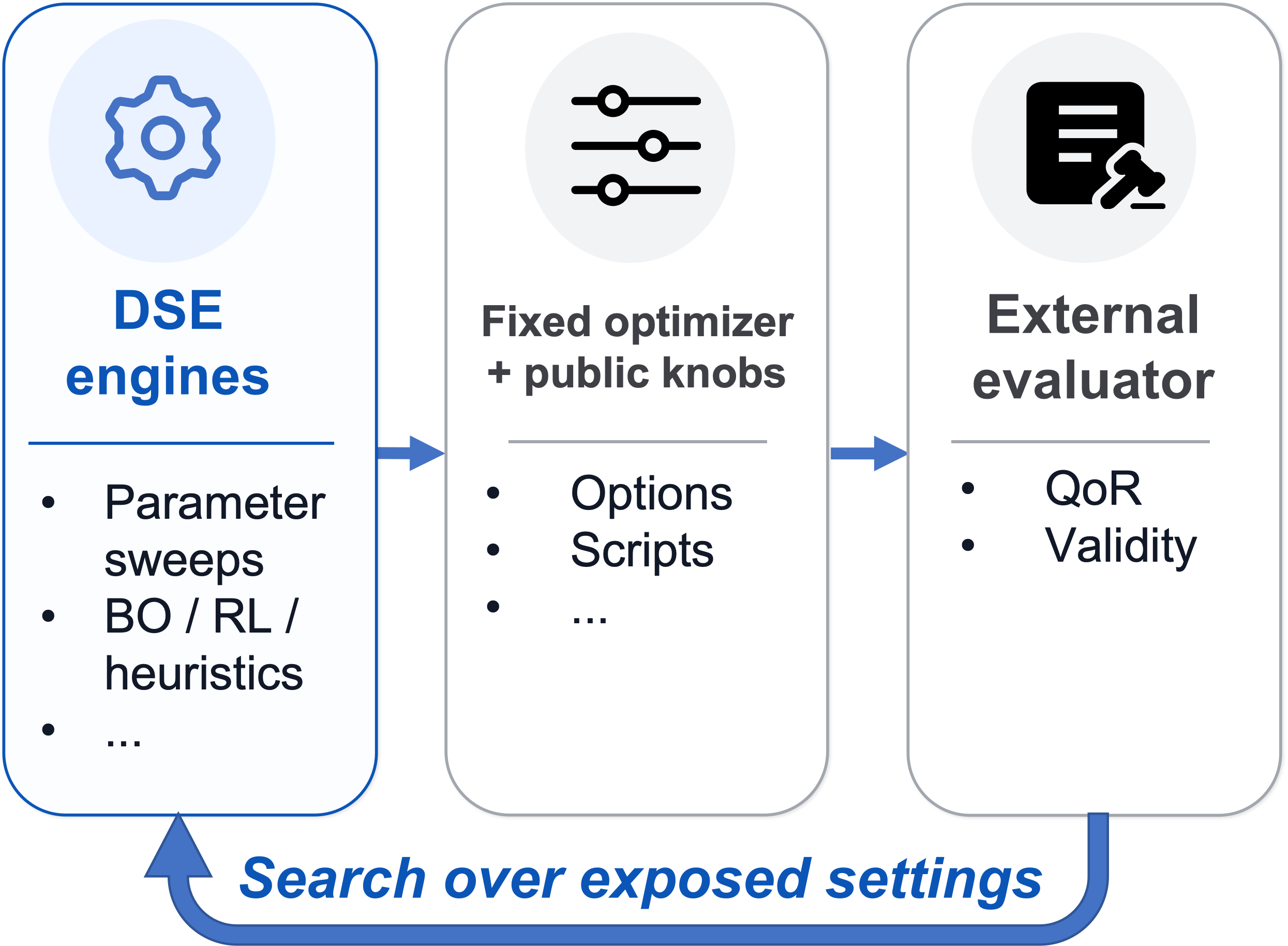}
    }
    \subfloat[White-box DSE]{
        \centering
        \label{subfig:black}
\includegraphics[width=.23\textwidth]{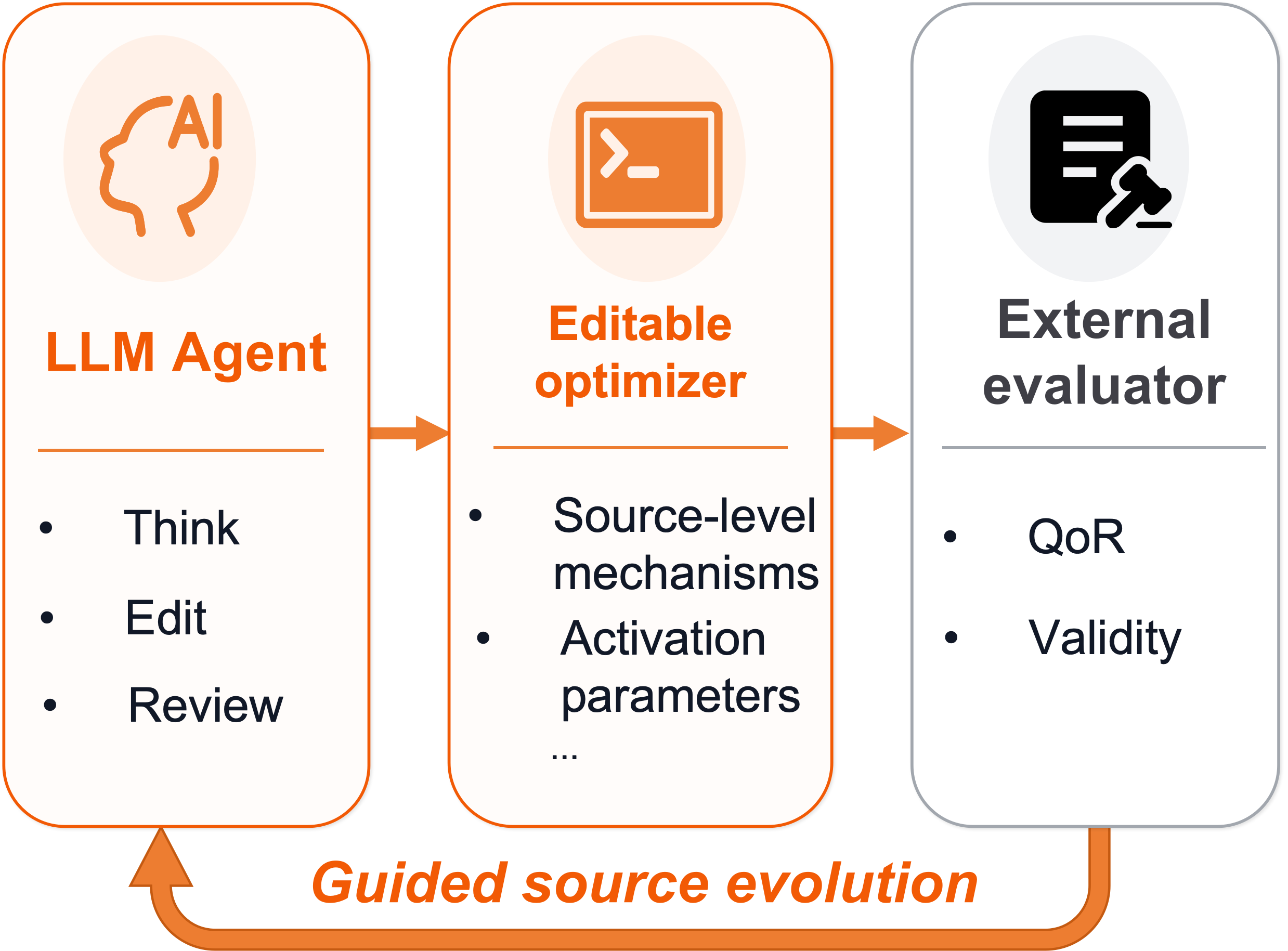}
    }
    \caption{{Search boundary for one target: black-box DSE tunes public knobs; white-box DSE \zyz{explores bounded} internal mechanisms under the protected evaluator.}}
    \Description{Two side-by-side diagrams contrast conventional black-box DSE, which tunes public tool controls, with white-box DSE, which explores bounded internal source mechanisms while preserving the evaluator.}
    \label{fig:whitebox-dse-boundary}
    \vspace{-0.6em}
\end{figure}


Open-source EDA changes where the DSE boundary can be drawn. In an open codebase such as OpenROAD~\cite{openroad}, an optimizer is not only invocable but also inspectable and modifiable. This enables white-box DSE for open-source EDA: instead of searching only over public knobs of a fixed executable, the search can enter selected parts of the optimizer implementation and explore bounded source-level mechanisms. In EDA flows, this mechanism space is naturally organized by stages and stage interfaces. A placement flow, for example, may include legalization policies, local improvement kernels, objective scoring, displacement control, rollback logic, and state passed from legalization to detailed-placement optimization. White-box DSE therefore asks which internal mechanisms should be changed, where they should be inserted, and how they should be configured for a target design.

The expanded mechanism space is difficult to explore manually. Source-level choices are numerous, implementation-dependent, and strongly coupled across stages. A local improvement in one stage may not survive the downstream flow; for example, a legalizer that reduces post-legalization HPWL can still produce a placement that is harder for detailed-placement optimization (DPO) to improve. Recent coding-evolution systems suggest a possible path forward: language-model agents can modify programs using evaluator feedback~\cite{novikov2025alphaevolve,assumpccao2025codeevolve,yu2025satlution}, and related ideas have begun to appear in EDA source-level evolution~\cite{yao2026evoplace,jafri2026grevolve,yu2026evoABC,amur2026AuDoPEDA}. These systems show that agents can generate useful code changes for complex tools.

\zyz{Source modification} alone does not make a reliable DSE procedure. Existing loops often treat each generated diff as a complete variant to be selected or rejected as a whole. This whole-variant view gives limited reusable evidence about which mechanism caused a gain, which stage activated it, whether it was live, whether it was runtime-expensive, or why it failed to compose with downstream optimization. Unconstrained source editing can also corrupt the evaluator path, change scripts or parsers, bypass legality checks, or produce stale source/binary/metric mismatches. Reliable white-box DSE therefore needs both a protected evaluator and a review unit finer than the complete diff.

We present ReviewDSE, an agentic white-box DSE framework for open-source EDA. ReviewDSE evaluates complete source candidates under the protected evaluator, while recording reusable search knowledge as reviewed stage/interface mechanism evidence. This evidence captures source references, liveness signals, compatibility observations, and promotion or rejection decisions. ReviewDSE uses a Teacher--Student agent structure: the Teacher reviews full-flow behavior and mechanism evidence, while Student agents adapt source starts, borrow peer mechanisms, or repair failed edits.

ReviewDSE follows a two-level workflow. Level~1 builds method evidence and source-start branches from calibration designs. Level~2 conducts target-design exploration through initialization based on the frozen evidence from Level~1, followed by iterative search driven by target-specific metrics, logs, diffs, peer candidates, and Teacher review. We instantiate ReviewDSE on OpenROAD detailed placement, where legalization, DPO, and their handoff expose strict legality and stage-composability trade-offs.

\noindent This paper makes the following major contributions:
\begin{itemize}[noitemsep,topsep=0pt,leftmargin=*]
    \item We formulate protected white-box DSE for open-source EDA, moving beyond public tool controls to bounded source-level mechanisms inside staged optimizers.

    \item We introduce ReviewDSE, a two-level source-mechanism exploration framework that builds calibration-time method evidence and source-start branches, then uses them to initialize target-specific exploration with Teacher review, liveness checks, full-flow validation, and target-local evidence.

    \item We instantiate ReviewDSE on OpenROAD detailed placement and show that reviewed source-mechanism exploration improves the post-DPL HPWL/runtime frontier beyond public-knob BO-DSE, exposes why stage-local winners require full-flow review, and recovers legality on hard cut-row stress tests.
\end{itemize}

\noindent The rest of the paper is organized as follows. Section~\ref{sec:background} reviews related work. Section~\ref{sec:methodology} presents ReviewDSE. Section~\ref{sec:experiments} reports the results. Sections~\ref{sec:discussion} and~\ref{sec:conclusion} discuss limitations and conclude the paper.

\section{Background and Related Work}
\label{sec:background}

\subsection{DSE in EDA}

Design space exploration (DSE) is widely used in EDA because QoR is highly sensitive to algorithm choices, tool parameters, and flow-stage interactions~\cite{geng2022edaDSEsurvey}. Prior work has explored DSE at multiple abstraction levels. In high-level synthesis, AutoDSE~\cite{sohrabizadeh2022autodse} searches HLS pragmas and transformations through bottleneck-guided exploration. In logic synthesis, BOiLS~\cite{grosnit2022boils} formulates synthesis-flow selection as a sequential optimization problem and searches logic-optimization recipes with Bayesian optimization. In physical design, placement-tuning methods optimize placement parameters and objectives using learning-based or multi-objective search~\cite{agnesina2020vlsi,agnesina2023autodmp}. These works show that automated search can improve QoR and reduce manual effort across EDA flows.

However, most existing EDA DSE methods remain black-box with respect to the optimizer implementation. Their search spaces are defined by exposed pragmas, parameters, command sequences, scripts, seeds, or flow-level configurations, while the tool source remains fixed. Such methods can improve how tools are invoked, but cannot directly explore internal mechanisms such as hidden cost models, repair policies, rollback decisions, or stage interfaces. Open-source infrastructures such as OpenROAD~\cite{openroad} make these implementation-level decisions searchable under the same external flow metrics, motivating white-box DSE where bounded source-level mechanisms become design choices.

\subsection{LLM Agents for EDA}
\label{sec:llm-eda}

Large language models have recently been explored in EDA as natural-language interfaces, code generators, and tool-assistance agents. One major line of work studies RTL or Verilog generation from natural-language specifications, with benchmarks such as RTLLM~\cite{lu2023rtllm} and VerilogEval~\cite{liu2023verilogeval} evaluating syntax, functional correctness, and design quality, and domain-specific models such as VeriGen~\cite{thakur2023verigen} and RTLCoder~\cite{liu2023rtlcoder} improving HDL generation through EDA-specific training. Beyond direct code generation, feedback-driven frameworks use compiler or simulation outputs to repair generated HDL, as in AutoChip~\cite{thakur2023autochip}. LLMs have also been used for hardware verification and testbench generation~\cite{zhang2023llm4dv,qiu2024autobench}. Another line of work uses LLMs as EDA tool assistants: ChatEDA~\cite{he2023chateda} supports task decomposition, script generation, and tool execution, ChipNeMo~\cite{liu2023chipnemo} studies domain-adapted assistants for chip-design tasks, and documentation-grounded systems answer questions over EDA manuals and OpenROAD documentation~\cite{pu2024rag,shi2024askeda}. These works improve how designers generate artifacts, query documentation, and invoke EDA tools, but they mostly operate around a fixed executable.

Recent coding-evolution systems show that LLM agents can modify programs using evaluator feedback~\cite{novikov2025alphaevolve,assumpccao2025codeevolve,yu2025satlution}. Similar ideas have begun to appear in EDA source-level evolution~\cite{yao2026evoplace,jafri2026grevolve,yu2026evoABC,amur2026AuDoPEDA}. In this setting, the LLM action moves from scripts or artifacts around the tool to the implementation of the tool itself. Self-Evolved ABC~\cite{yu2026evoABC} evolves a logic-synthesis codebase under correctness and synthesis-QoR feedback. AuDoPEDA~\cite{amur2026AuDoPEDA} produces repository-grounded OpenROAD diffs using code-graph documentation and QoR-driven execution. GR-Evolve~\cite{jafri2026grevolve} specializes global-router source code to individual designs, and EvoPlace~\cite{yao2026evoplace} evolves global-placement optimization components.

These works establish that LLM agents can improve EDA tools through source modification. ReviewDSE addresses a complementary question: how to make source modification a reliable white-box DSE procedure. Rather than treating each generated diff only as a whole-program winner or loser, ReviewDSE evaluates complete candidates under a protected flow while recording reusable positive and negative evidence at the mechanism level. This allows useful calibration evidence, stage donors, and rejection decisions to guide later search rounds. Table~\ref{tab:related-position} summarizes the position of ReviewDSE relative to prior directions. ReviewDSE still evaluates complete source candidates, but its reusable memory is mechanism-level evidence that later agents can refine or combine during white-box optimization.

\begin{table}[tb]
    \centering
    \caption{{Positioning of ReviewDSE.}}
    \label{tab:related-position}
    \scriptsize
    \setlength{\tabcolsep}{2pt}
    \resizebox{\columnwidth}{!}{%
    \begin{tabular}{@{}llll@{}}
        \toprule
        Category & Search point & Review signal & Search memory \\
        \midrule
        EDA parameter DSE & knobs/scripts & QoR logs & best settings \\
        LLM tool use & artifacts/scripts & tool logs & revised artifact \\
        Source evolution & source patch & build/QoR logs & winning patch \\
        ReviewDSE & source mechanism edit & protected full-flow evidence & reusable mechanism evidence \\
        \bottomrule
    \end{tabular}
    }
\end{table}

\section{Methodology}
\label{sec:methodology}

\begin{figure*}[tb]
    \centering
    \includegraphics[width=0.95\linewidth]{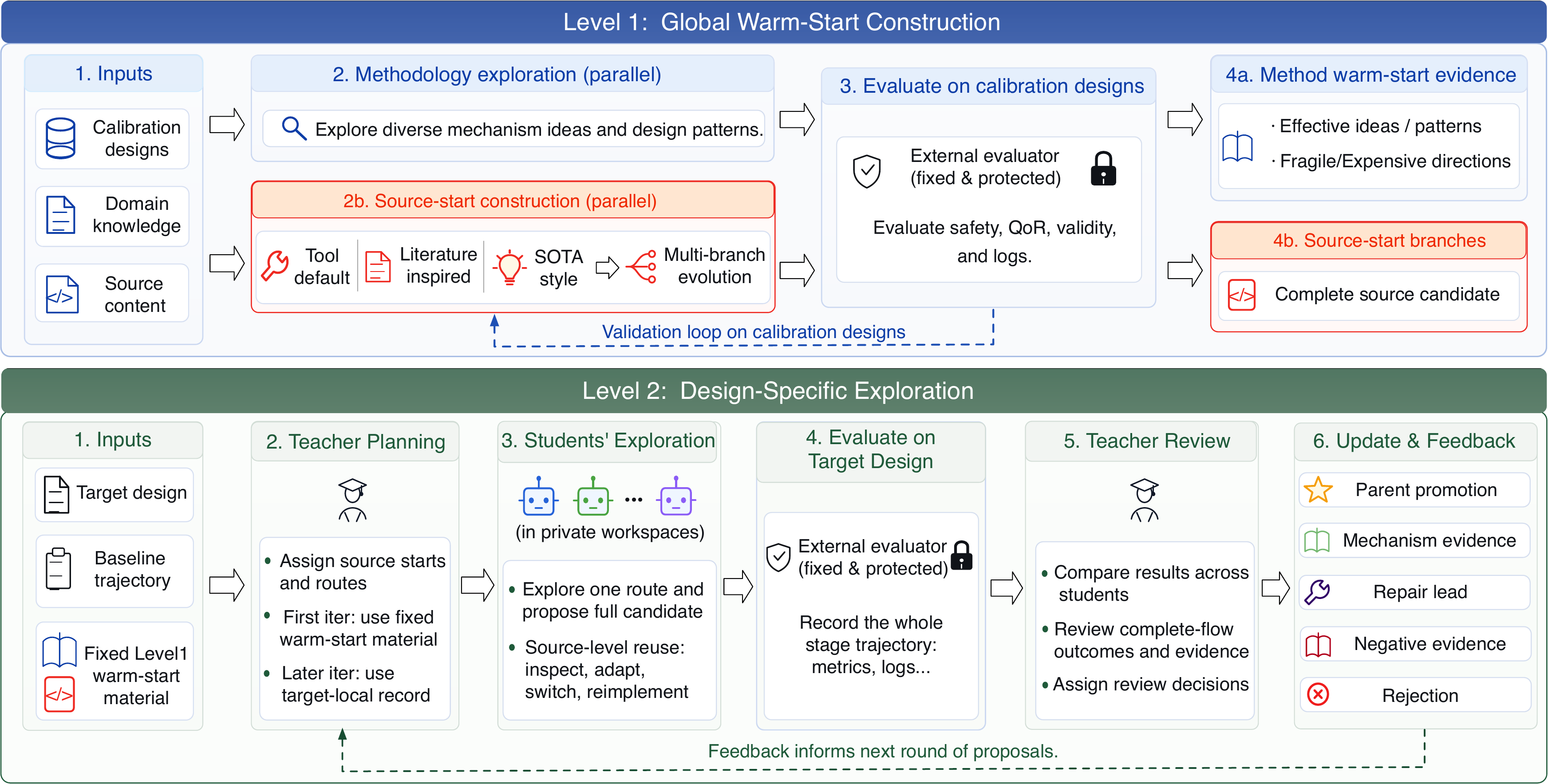}
    \caption{ReviewDSE workflow. Level~1 builds fixed warm-start evidence and source-start branches. Level~2 initializes target search from them, then \zyz{explores} candidates under the protected evaluator.}
    \Description{Workflow diagram showing Level 1 calibration producing reviewed mechanism evidence and source-start branches, followed by Level 2 target search with Teacher review, Student edits, and protected full-flow evaluation.}
    \label{fig:workflow}
\end{figure*}

\subsection{Formulation and Overview}
\label{sec:formulation}

Conventional black-box DSE treats the EDA optimizer as a fixed executable evaluated by an external flow-level evaluator. Let $O$ denote the implementation of a staged optimizer for a target design $d$, let $E$ denote the protected evaluator used to run the flow and collect canonical metrics, and let $x \in \mathcal{X}$ denote the public configuration vector exposed through command options, flow scripts, pass sequences, or other tool parameters. Black-box DSE searches over $x \in \mathcal{X}$. Each search point is evaluated as $E(O,x,d)$, while the optimizer implementation $O$ and the measurement path $E$ remain fixed. This formulation is practical, but it limits the search to behaviors already exposed by the tool developer.

ReviewDSE keeps the same evaluator $E$ but expands the search variable from a public configuration to a bounded source edit. Let $\mathcal{S}$ denote the editable optimizer source surface and let $\mathcal{D}(\mathcal{S})$ denote the admissible edits on that surface. ReviewDSE searches over $\Delta \in \mathcal{D}(\mathcal{S})$. Each search point produces a complete source candidate $O+\Delta$, which is built and evaluated on $d$ as $E(O+\Delta,d)$.

EDA optimizers are naturally organized as staged flows, with intermediate states passed across neighboring stages. ReviewDSE uses this native structure to define review scopes for source edits. Let $\mathcal{C}$ denote the set of stage and interface scopes, where each $c \in \mathcal{C}$ corresponds to either an existing optimization stage or a handoff between two neighboring stages. Each scope has a pool of candidate mechanisms,
\begin{equation}
\mathcal{M}_c = \{p_{c,1}, \ldots, p_{c,n_c}\}.
\end{equation}
A mechanism is not the stage itself. It is a source-level logic chain that changes how the stage initializes state, generates candidates, scores moves, accepts or rolls back transactions, exports handoff state, or performs local repair. A mechanism may also be controlled by activation parameters $\theta_{c,j}$, such as weights, thresholds, budgets, seeds, fallback rules, or early-stop conditions.

A candidate edit $\Delta$ may touch one or more native scopes. For each scope $c$, let $\mathcal{M}_c(\Delta) \subseteq \mathcal{M}_c$ be the mechanisms implemented by the source candidate produced by $\Delta$. The mechanism content of the candidate can be written as
\begin{equation}
\mathsf{mech}(\Delta) =
\bigcup_{c \in \mathcal{C}}
\{(c,p_{c,j},\theta_{c,j}) \mid p_{c,j} \in \mathcal{M}_c(\Delta)\}.
\end{equation}
This notation describes the mechanisms realized by an implemented source candidate. It does not imply that ReviewDSE materializes standalone mechanism units and assembles them into a tool variant. The actual search point is a source workspace built and evaluated as $O+\Delta$. After evaluation, ReviewDSE retains mechanism-level evidence rather than reusable code objects. As summarized in~\Cref{tab:unit}, this evidence records source anchors, liveness signals, full-flow behavior, and review outcomes, allowing later search to inspect, adapt, or reimplement the reviewed mechanism logic in another candidate. A record may indicate that the mechanism improves the target, is active but harmful, never executes, depends on a compatible handoff, or should be avoided.


\begin{table}[tb]
    \centering
    \caption{{Reviewed mechanism evidence record.}}
    \label{tab:unit}
    \footnotesize
    \renewcommand{\arraystretch}{1.02}
    \begin{tabular}{p{0.28\linewidth}p{0.64\linewidth}}
        \hline
        \textbf{Field} & \textbf{Recorded information} \\
        \hline
        {Scope} & {Target stage or stage interface, used to relate evidence to stage-wise behavior.} \\
        {Source reference} & {Branch or patch reference, modified files, code anchors, and mechanism intent.} \\
        {Controls/liveness} & {Activation parameters, counters, traces, and artifacts showing execution.} \\
        {Review outcome} & {Full-flow metrics, compatibility observations, and Teacher decisions such as parent promotion, mechanism evidence, repair lead, negative evidence, or rejection.} \\
        \hline
    \end{tabular}
\end{table}


\subsection{Protected Evaluation Contract}
\label{sec:protected-evaluator}

The protected evaluator is the trust boundary for source-level DSE. It fixes the benchmark inputs, incoming design state, top-level command sequence, baseline harness, legality or validity checker, metric parser, reference comparison, runtime gate, evaluator scripts, and unrelated tool components. Evaluator-produced objective and stage metrics remain outside the editable source surface. A candidate is rejected if it changes or bypasses this path, fails to build, violates validity checks, misses canonical metrics, exceeds the runtime gate, or lacks liveness evidence for its claimed mechanism.

Figure~\ref{fig:workflow} summarizes how ReviewDSE works. Level~1 constructs frozen warm-start evidence before target optimization: method evidence records calibration-observed mechanism routes, and source-start construction produces complete source basins. Level~2 uses this evidence as prior guidance, but every promoted candidate is rebuilt, executed, and reviewed on the current target case. Level~2 may accumulate target-local evidence, but it does not update the global warm-start evidence.

\begin{table}[tb]
\centering
\caption{{Teacher review decisions used in \zyz{exploration}.}}
\label{tab:review-outcomes}
\scriptsize
\setlength{\tabcolsep}{2pt}
\begin{tabular}{@{}p{0.25\linewidth}p{0.48\linewidth}p{0.19\linewidth}@{}}
\toprule
Decision & Condition & Use \\
\midrule
Parent promotion & Valid, complete metrics, live, within gate, improves target score & Next parent \\
Mechanism evidence & Active mechanism helps a stage or interface but loses final score & Evidence record \\
Repair lead & Explains validity/runtime failure or fragile handoff & Targeted route \\
Negative evidence & Inactive, too expensive, or non-composable & Search constraint \\
Candidate rejection & Build, validity, metric, stale-binary, or evaluator-path failure & Discard \\
\bottomrule
\end{tabular}
\end{table}

\subsection{Calibration Warm-Start Construction}
\label{sec:warm-starts}
Level~1 prepares the search by building two fixed warm-start products before any target case is optimized: method warm-start and source-start branches. It does so by using calibration designs as diagnostic probes. These designs are not target designs. Instead, they are constructed to expose different optimizer behaviors, including stage-local damage, downstream recoverability, constraint pressure, runtime pressure, and stage-interface incompatibility. Level~1 converts these observations into mechanism hypotheses and executable starting points, so that Level~2 does not begin from a blank prompt.

Method warm-start construction focuses on the hypothesis side. Agents inspect papers, source code, and domain knowledge, then implement candidate mechanisms on existing branches and evaluate them with the protected flow. Level~1 review summarizes the calibration behavior of each idea, including whether it is promising, fragile, too expensive, inactive, or non-composable. The resulting records provide route and diagnosis evidence for the Level~2 Teacher when a target case shows similar stage movement or design characteristics.

Source-start construction focuses on the implementation side. Some literature-inspired or SOTA-style ideas are too large to recreate in each target iteration, so Level~1 materializes them as complete source branches during calibration. In Level~2, these branches serve as editable starts rather than certified target solutions. Students may extend one, switch to another, or reimplement a useful logic chain after inspecting peer diffs. This gives the target search concrete code for implementing or repairing a route suggested by the method evidence.


\subsection{Target-Case Source Exploration Loop}
\label{sec:design-specific-co-evolution}

Level~2 runs an independent white-box optimization loop for each target design. For a new target, ReviewDSE first measures the target baseline trajectory and creates private Student workspaces. The loop maintains a local record that includes baseline metrics, evaluated candidates, stable source references, implementation diffs, Student knowledge cards, Teacher reviews, peer-learning packets, and any clean promoted parent. The Level~1 warm-start material is only used as prior knowledge, and subsequent decisions are driven by the local record accumulated on the current design.

In the first iteration, the Teacher compares the target baseline trajectory and design characteristics with the fixed Level~1 method evidence and available source branches. Based on this comparison, the Teacher assigns each Student an initial source start and mechanism route, with generated packets and route tables as supporting evidence. This assignment only defines the starting direction for the target search. The selected branch is not fixed across iterations. Later rounds may extend it if it remains effective, or switch to repairs, alternative branches, or new mechanisms when target-local evidence suggests a better route.

In later iterations, the Teacher first reviews the previous round and then decides how to balance continuation and exploration. A clean candidate with the best final objective inside the runtime gate may seed an elite continuation branch, typically for one Student. The other Students can explore diversity routes or reuse evidence from peer candidates. This reuse remains source-level: a Student may inspect another candidate's diff, summarize its logic chain, adapt a patch, switch to a branch, or reimplement the idea in its own workspace. The reviewed mechanism guides the edit, but it is not copied as an independent plug-in.

Each Student edits a private optimizer source workspace and submits a complete $O+\Delta$ candidate. The Student must build or relink a private binary, run the fixed evaluator with that binary, and provide complete artifacts. Because earlier-stage choices affect later stages, the Student is encouraged to reason about the whole stage trajectory rather than optimizing a stage metric in isolation.

After evaluation, the Teacher reviews the complete-flow outcome together with the mechanism evidence behind it, and assigns a review decision following~\Cref{tab:review-outcomes}. A valid improved candidate may become the promoted parent for later target search. Other evaluated candidates may still be retained as target-local evidence when their source diffs and logs explain an active mechanism, even if they are not promoted. Invalid candidates are recorded but cannot be promoted. These review outcomes update only the target-local record, while the Level~1 warm-start evidence remains unchanged.

\subsection{Instantiation on OpenROAD Detailed Placement}
\label{sec:openroad-dpl-instantiation}

\begin{figure}[tb]
    \centering
    \includegraphics[width=0.98\linewidth]{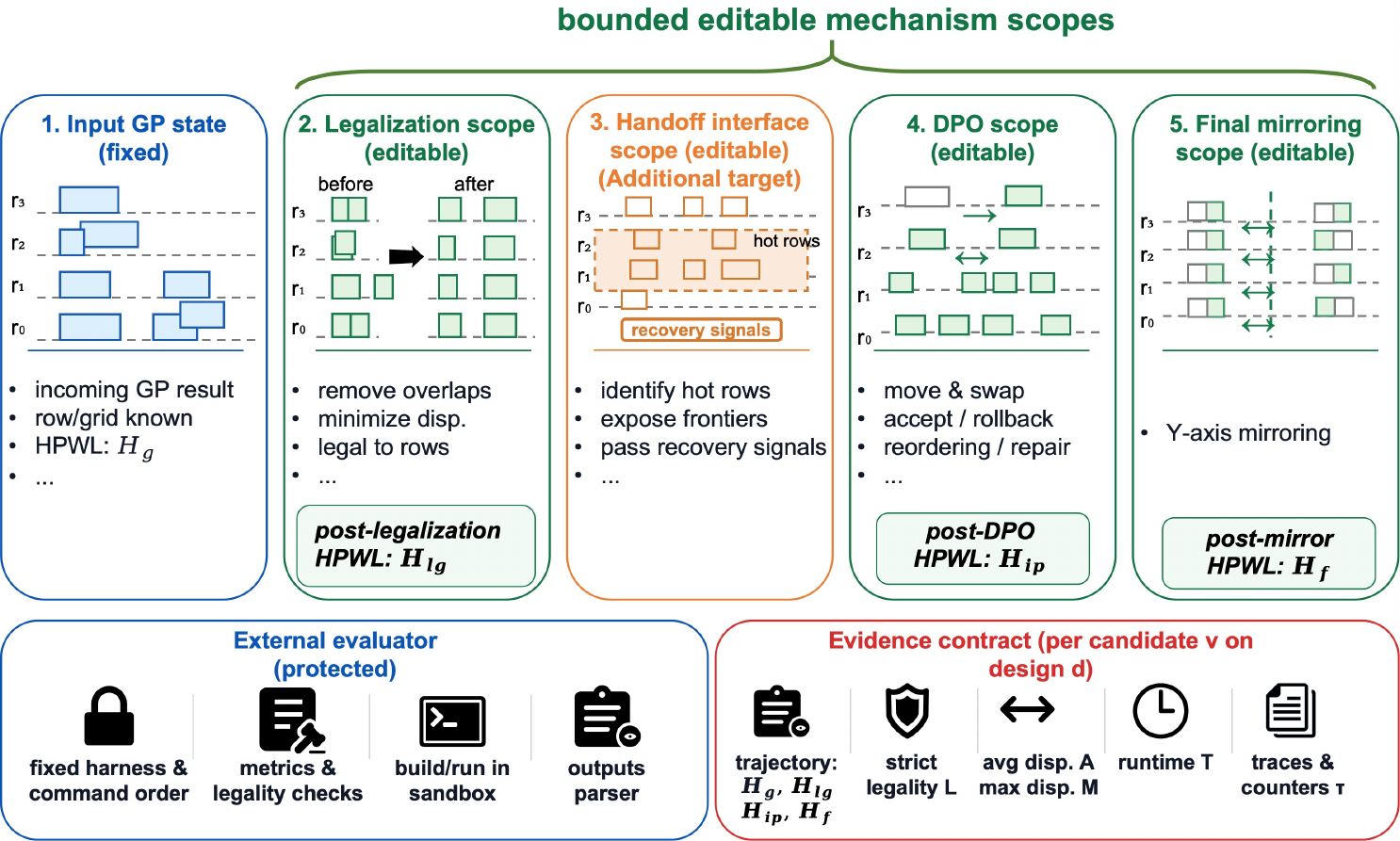}
    \caption{OpenROAD DPL instantiation. ReviewDSE edits bounded legalization, DPO, and handoff mechanisms while keeping the evaluation path protected.}
    \Description{Block diagram of the OpenROAD detailed-placement instantiation, separating editable legalization, detailed-placement optimization, and handoff mechanisms from the protected build, execution, legality, and metric path.}
    \label{fig:dpl-evolve-casestudy}
\end{figure}

We instantiate ReviewDSE on OpenROAD detailed placement.\footnote{We use constructed Nangate45 detailed-placement calibration cases for Level~1: JPEG at UTIL=90, AES at UTIL=70, and a SWERV wrapper at UTIL=60.} We call the \zyz{source-specialized} commands DPL-evolve, DPO-evolve, and mirror-evolve.
Figure~\ref{fig:dpl-evolve-casestudy} summarizes the editable and protected parts of this instantiation.
The evaluated command sequence is fixed and keeps the canonical three stages: legalization, DPO, and mirroring.
ReviewDSE edits only the bounded source implementations invoked by these stages in private workspaces.
The strict comparison anchor is the fixed OpenROAD detailed-placement flow.

Following the formulation above, we use legalization and DPO as the two main stage scopes, and we use the legalization-DPO handoff as the main interface scope.
In legalization, source-start branches include Diamond-style, Negotiation-based, and LEGALM-inspired methods.
Within this scope, ReviewDSE may explore row-assignment policies, row-segment reservation, overflow-aware relocation, connectivity-aware escape, bounded compaction, and post-legality repair.

DPO \zyz{exploration} starts from the same native improve-placement source inside the selected legalizer branch.
The Teacher assigns the mechanism implementation path, and the Student edits the relevant DPO source path.
The native improve-placement behavior contains local kernels such as independent-set matching, global swaps, vertical swaps, row-local reordering, randomized local improvement, and legality or orientation checks.
ReviewDSE may revise or extend this source with local-move generation, partner selection, region kicks, dirty-window dynamic programming, exact affected-net HPWL scoring, transaction accept/rollback, fallback descent, per-net extrema caches, and thread-local scratch storage.
Handoff mechanisms include dirty rows, pressure regions, touched-net frontiers, boundary-active cells, local repair windows, and source-produced state that downstream DPO can consume.

For an evaluated candidate $O+\Delta$ on design $d$, the evaluator records the full stage trajectory
\begin{equation}
E(O+\Delta,d) = \{H_g, H_{lg}, H_{ip}, H_f, L, A, M, T, \tau\},
\end{equation}
where $H_g$, $H_{lg}$, $H_{ip}$, and $H_f$ are half-perimeter wirelength (HPWL) after incoming global placement, legalization, DPO, and final post-DPL evaluation.
$L$ denotes strict legality, $A$ and $M$ are average and maximum displacement, $T$ is runtime, and $\tau$ contains traces, liveness counters, and diagnostics.
Full-trajectory review is necessary because a stage-local improvement is not always a valid target improvement.
A lower legalization HPWL $H_{lg}$ may still worsen the final post-DPL HPWL $H_f$ after DPO and mirroring.
Conversely, a non-winning candidate can also provide active mechanism evidence under specific target patterns.
This is why ReviewDSE retains mechanism evidence for later target search.

\section{Experiments}
\label{sec:experiments}

Our experiments compare white-box source optimization with black-box DSE, test runtime-aware selection, examine full-flow composability, and evaluate whether \zyz{source-mechanism exploration} can recover legality on hard cut-row layouts.

\subsection{Experiment Setup and Metrics}
\label{sec:protocol}

{We use a fixed OpenROAD/\orfs{} checkout in the protected evaluator and run all methods through the same measurement path. The exact checkout, evaluator scripts, and selected-candidate artifacts are recorded with the experiment data.}
{The main target search uses one Teacher agent with the GPT-5.5 xhigh profile and four parallel Student agents with the GPT-5.4 xhigh profile for 10 review iterations.}
{We report only candidates that compile, pass strict legality, produce complete protected metrics, satisfy source/binary/metric consistency, show liveness, and stay within a \(2\times\) runtime gate.}

We compare against BO-DSE, a representative black-box baseline with fixed OpenROAD source.
BO-DSE uses Bayesian optimization (BO) with Optuna's Tree-structured Parzen Estimator (TPE) sampler.
It searches exposed legalization, DPO, extra-invocation, and global-swap controls for 400 trials per case with four-way parallelism.

Our experimental dataset comprises nine target tasks derived from the AES, Ariane133, Ibex, JPEG, and SWERV designs, implemented in ASAP7 and Nangate45 technology nodes. The calibration and target sets are disjoint evaluation instances. They may share benchmark families, but no target placement, target metric, target route, or target-specific candidate is used in Level~1. {All target designs are evaluated independently by the protected flow.}

To evaluate quality of results (QoR), we use the final post-DPL HPWL, denoted as \(\hf\), as the primary metric.
We report the relative HPWL change \(\Delta\hf\) and runtime compared with the default Diamond detailed-placement flow.
For ReviewDSE, each test case reports both the best-\(\hf\) candidate that satisfies the \(2\times\) runtime gate and a lower-overhead candidate selected by a runtime-aware score \(G_{HR}\)\footnote{The higher-is-better selection score is defined as \(G_{HR}=100\cdot(\hf^{Default}-\hf)/\hf^{Default}-P(T/T^{Default})\), where \(T\) is the runtime, \(P(r)=0\) for \(r\le1.10\), and \(P(r)=(\sqrt{r}-\sqrt{1.10})/(\sqrt{2}-\sqrt{1.10})\) for \(r>1.10\). A \(2\times\) runtime penalty requires a one percentage point reduction in post-DPL HPWL to break even; small runtime noise is ignored.}.

\subsection{{White-Box Source Optimization}}

\begin{table}[tb]
  \centering
  \caption{Nine-case DSE results.}
  \label{tab:dse-comparison}
  \scriptsize
  \setlength{\tabcolsep}{1.5pt}
  \resizebox{0.98\columnwidth}{!}{%
  \begin{tabular}{@{}lccccc@{}}
    \toprule
    Design & Default & BO & ReviewDSE-HPWL & ReviewDSE-\(G_{HR}\) & Tokens \\
    \midrule
    AES ASAP7 & 42.2k/5.3s & -0.98\%/1.47\(\times\) & -1.91\%/1.91\(\times\) i10 & -1.40\%/1.38\(\times\) i10 & 1.62/0.08 \\
    AES N45 & 176.8k/6.0s & -1.20\%/1.38\(\times\) & -3.67\%/1.86\(\times\) i10 & -3.67\%/1.86\(\times\) i10 & 1.81/0.09 \\
    Ariane133 N45 & 5.78M/151s & -0.06\%/0.92\(\times\) & -5.50\%/0.56\(\times\) i10 & -5.50\%/0.56\(\times\) i10 & 3.66/0.11 \\
    Ibex ASAP7 & 63.3k/6.9s & -0.10\%/1.10\(\times\) & -0.60\%/1.17\(\times\) i10 & -0.60\%/1.17\(\times\) i10 & 1.68/0.10 \\
    Ibex N45 & 200.1k/5.1s & -0.09\%/1.00\(\times\) & -0.99\%/2.00\(\times\) i10 & -0.56\%/1.08\(\times\) i6 & 1.52/0.12 \\
    JPEG ASAP7 & 134.6k/23s & -0.63\%/1.39\(\times\) & -0.94\%/1.00\(\times\) i5 & -0.94\%/1.02\(\times\) i3 & 1.87/0.10 \\
    JPEG N45 & 467.2k/33s & -0.23\%/1.13\(\times\) & -1.63\%/0.56\(\times\) i9 & -1.63\%/0.77\(\times\) i10 & 1.93/0.09 \\
    SWERV ASAP7 & 913.9k/57s & -0.04\%/1.08\(\times\) & -0.33\%/1.90\(\times\) i8 & -0.30\%/1.09\(\times\) i4 & 3.06/0.09 \\
    SWERV N45 & 3.27M/34s & -0.09\%/1.08\(\times\) & -0.49\%/1.07\(\times\) i9 & -0.49\%/1.07\(\times\) i9 & 2.14/0.14 \\
    \midrule
    Mean & -- & -0.38\%/1.17\(\times\) & -1.78\%/1.34\(\times\) & -1.68\%/1.11\(\times\) & 2.15/0.10 \\
    \bottomrule
  \end{tabular}
  }
  \vspace{1pt}
  \footnotesize Entries are post-DPL HPWL delta/runtime vs. default; \(i\#\) is the selected iteration. Tokens are logged/active budgets in billions.
\end{table}

\begin{figure}[tb]
    \centering
    \includegraphics[width=0.90\linewidth]{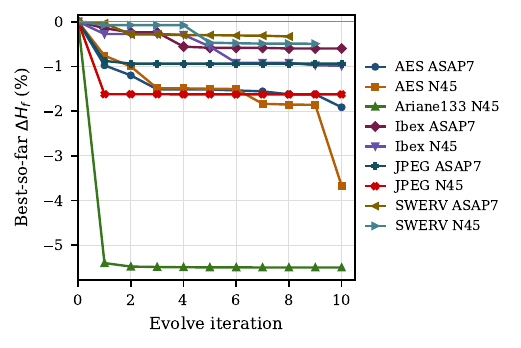}
    \caption{{10-iteration best-so-far \(\Delta H_f\).}}
    \Description{Nine best-so-far curves across ten iterations show post-DPL HPWL reductions for the evaluated target cases; downward steps indicate newly promoted improvements.}
    \label{fig:evolve-hpwl-curves}
\end{figure}

{Table~\ref{tab:dse-comparison} compares the OpenROAD default, public-knob BO-DSE, and ReviewDSE on the nine target tasks. After 400 trials per case, BO-DSE improves mean post-DPL HPWL by 0.38\%, while ReviewDSE improves all nine targets by 1.78\% on average within the same \(2\times\) runtime gate. Runtime-aware ReviewDSE-\(G_{HR}\) preserves a 1.68\% mean HPWL reduction while reducing average runtime from 1.34\(\times\) to 1.11\(\times\). This comparison uses the same protected evaluator and target tasks, but the two methods search different spaces: BO-DSE searches exposed controls of a fixed source tree, while ReviewDSE explores reviewed source mechanisms.
Calibration evidence only initializes promising source directions. Every reported candidate is rebuilt, evaluated, and re-validated on its own target by the protected evaluator. The token column reports mean per-target search cost.}

{\Cref{fig:evolve-hpwl-curves} shows best-so-far \(\Delta\hf\) over 10 target iterations. Early drops indicate that Level~1 evidence can choose useful initial source directions.\footnote{~In tested Ariane133 N45 diagnostic runs, omitting Diamond-sourceTopK averaged \(+1.52\%\)/\(1.07\times\); matching Level~1 guidance selected it in iteration~1 and averaged \(-3.26\%\)/\(0.71\times\), yielding a local parent. Diamond-sourceTopK scores top-K DPO moves/swaps from a Diamond basin and passes hot nodes/segments to reorder. This is not a separate evaluation target.} Later drops, such as AES N45, reflect target-specific source exploration after Teacher review, runtime feedback, and peer-candidate evidence. Flat segments correspond to iterations without a promoted parent, often because review rejects non-composable or runtime-expensive candidates.}

\begin{figure}[tb]
    \centering
    \includegraphics[width=0.98\linewidth]{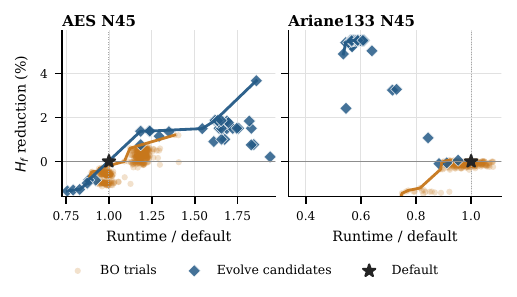}
    \caption{{Runtime--quality frontier on two Nangate45 cases; \(y\)-axis is post-DPL HPWL reduction vs. default.}}
    \Description{Two scatter plots compare normalized runtime and post-DPL HPWL reduction for BO-DSE trials and ReviewDSE candidates on AES and Ariane133 Nangate45 cases, with the default point shown for reference.}
    \label{fig:bo-evolve-pareto}
\end{figure}

{\Cref{fig:bo-evolve-pareto} plots HPWL reduction against normalized runtime for two Nangate45 cases. On AES N45, ReviewDSE reaches a 3.67\% HPWL reduction and outperforms the best BO-DSE point at comparable runtime. On Ariane133 N45, ReviewDSE reaches a 5.50\% HPWL reduction while reducing runtime relative to the default. These tradeoffs were not observed in the 400-trial BO-DSE search over the specified public-knob space.}

\subsection{{Full-Flow Review and Downstream Composability}}
{We next examine why ReviewDSE reviews complete trajectories instead of accepting stage-local winners. The test applies a deliberately stage-local selection rule: it selects candidate legalizers by post-legalization HPWL \(H_{lg}\), then passes the selected placements through the same downstream DPO and final mirroring flow. The dense N45 cases are constructed validation variants with increased legalization pressure and are not included in the nine-case QoR average. Table~\ref{tab:stage-line-full} reports cases where the selected legalizer improves \(H_{lg}\) but degrades the final post-DPL HPWL \(H_f\).}

\begin{table}[tb]
\centering
\caption{Stage-local legalization counterexamples.}
\label{tab:stage-line-full}
\scriptsize
\setlength{\tabcolsep}{2pt}
\resizebox{\columnwidth}{!}{%
\begin{tabular}{@{}lllrrrr@{}}
\toprule
Case & Selected legalizer & Full-flow reference & \(H_{lg}\) / ref. & \(\Delta H_{lg}\) & \(H_f\) / ref. & \(\Delta H_f\) \\
\midrule
AES dense N45 & LEGALM & Diamond & 137.8k / 138.8k & -0.76\% & 130.8k / 128.9k & +1.48\% \\
JPEG dense N45 & Negotiation & Negotiation & 1.26M / 1.48M & -14.96\% & 711.7k / 588.3k & +20.96\% \\
SWERV dense N45 & Diamond & Negotiation & 2.81M / 2.81M & -0.12\% & 2.71M / 2.71M & +0.02\% \\
\bottomrule
\end{tabular}
}
\end{table}

These counterexamples show that a lower-HPWL legalized placement can be less compatible with downstream DPO. JPEG dense N45 is the clearest case: the selected legalizer reduces \(H_{lg}\) by 14.96\%, but the final \(H_f\) worsens by 20.96\% after DPO and mirroring. Thus, legalization quality alone is not a sufficient promotion criterion.

\subsection{Hard Cut-Row Repair}
Finally, we test legality recovery on hard cut-row stress tests. These layouts contain fragmented row topologies that cause fixed-source reference routes to fail legality, time out, or miss a strict legal solution. Unlike the nine-case QoR comparison, the objective here is to recover legal placements under the same protected legality checker and runtime cap.

\begin{table}[tb]
\centering
\caption{Pattern-level hard cut-row repair evidence.}
\label{tab:cutrow-repair-full}
\scriptsize
\setlength{\tabcolsep}{2pt}
{\renewcommand{\arraystretch}{0.92}
\resizebox{\columnwidth}{!}{%
\begin{tabular}{@{}llccl@{}}
\toprule
Case & Cut-Row Pattern & Fixed ref (D/N) & ReviewDSE & Evidence \\
\midrule
\multirow{3}{*}{Ariane133 N45}
 & center-8um  & fail/pass & pass, 42.2s & beats legal N by 1.03\%, 18.9\(\times\) \\
 & center-9um  & fail/fail & pass, 51.6s & strict repair; no fixed legal ref. \\
 & center-10um & fail/TO   & pass, 59.6s & strict repair; N reaches timeout \\
\midrule
\multirow{3}{*}{SWERV dense N45}
 & center-5.25um & pass/fail & pass, 48.6s & beats legal fixed ref. by 3.69\%, 3.4\(\times\) \\
 & center-5.5um  & fail/fail & pass, 50.8s & strict repair; no fixed legal ref. \\
 & center-6um    & fail/fail & pass, 51.7s & strict repair; no fixed legal ref. \\
\midrule
\multirow{3}{*}{BPQUAD}
 & center-5um & fail/TO & pass, 563.6s  & D fails; N reaches 7200s cap \\
 & center-6um & fail/TO & pass, 799.4s  & D fails; N reaches 7200s cap \\
 & center-8um & fail/TO & pass, 1085.0s & D fails; N reaches 7200s cap \\
\bottomrule
\end{tabular}
}
}
\vspace{1pt}
\footnotesize D/N = Diamond/Negotiation; TO = 7200s timeout. center-\(x\)um is a vertical center-band cut-row with \(x\)\(\mu\)m halo. pass means OpenROAD and \texttt{check\_placement} pass.
\end{table}

Table~\ref{tab:cutrow-repair-full} reports pattern-level replay evidence rather than only family-level counts.
ReviewDSE produces strict legal placements on all nine hard cut-row patterns.
For Ariane133 and SWERV, at least one pattern also has a legal fixed-source reference, and ReviewDSE improves both HPWL and runtime.
For BPQUAD, the fixed routes fail or time out, so the evidence is legality recovery within 563.6--1085.0s rather than QoR dominance.
These results complement Table~\ref{tab:dse-comparison} by showing that reviewed source-mechanism exploration can recover strict legality on failure-oriented stress tests, not only improve QoR on already feasible targets.

\section{Discussion}
\label{sec:discussion}

{ReviewDSE intentionally performs target-case source specialization, not universal optimizer replacement. Calibration provides method evidence and source starts, but each reported candidate is rebuilt, executed, and promoted only on its own target under the protected evaluator. The fixed measurement path prevents metric gaming, and promotion also requires canonical metrics, source/binary/metric consistency, runtime gating, and liveness evidence.}

{The main limitations are evaluator scope and search cost. Current results cover HPWL, runtime, displacement, and strict legality; timing, congestion, power, and signoff require additional protected metrics. The target searches average 2.15B logged tokens and 0.10B active uncached-input-plus-output tokens over 10 iterations, so ReviewDSE is most appropriate for high-value designs, hard failures, or cases where public knobs have limited remaining headroom.}

\section{Conclusion}
\label{sec:conclusion}

{ReviewDSE reframes open-source EDA tuning as protected white-box DSE: for a target design, bounded optimizer mechanisms can be specialized under a fixed evaluator rather than only invoked through public knobs. In OpenROAD detailed placement, this source-level search boundary improves the post-DPL HPWL/runtime frontier beyond public-knob BO-DSE, reveals why stage-local winners require full-flow review, and recovers legality on hard cut-row stress tests. The main lesson is that \zyz{source-mechanism exploration} for EDA should not be treated as unrestricted tool rewriting: complete candidates must be judged by canonical metrics and legality checks, while reusable knowledge is accumulated as reviewed mechanism evidence.}

\begin{acks}
The work is supported by 
the AI for Science Program, Shanghai Municipal Commission of Economy and Informatization (2025-GZL-RGZN-BTBX02038), 
and Fudan Kunpeng\&Ascend Center of Cultivation.
\end{acks}

\balance
\newpage
\nobalance
{
    \bibliographystyle{ACM-Reference-Format}
    \bibliography{ref/Top,ref/reference}
}

\newpage
\appendix
\section{Artifact Appendix}
\begingroup
\raggedbottom
\normalsize
\setlength{\emergencystretch}{2em}
\setlength{\parskip}{0pt}
\setlength{\textfloatsep}{3pt plus 1pt minus 1pt}
\setlength{\floatsep}{3pt plus 1pt minus 1pt}
\setlength{\intextsep}{3pt plus 1pt minus 1pt}
\setlength{\abovecaptionskip}{1pt}
\setlength{\belowcaptionskip}{1pt}

\noindent \href{mailto:25303060069@m.fudan.edu.cn}{Wenjie Yuan} of Fudan University maintains the artifact package and its reproduction workflow.

\subsection{Abstract}

Our artifact contains the ReviewDSE implementation, the prompts and protected
evaluator used in the experiments, and the source programs evaluated in the
paper. It also includes a pinned open-source EDA environment and scripts for
reproducing Tables~4--6 and Figures~4--5. The artifact runs on Rocky
Linux~8.10 (x86-64) and requires authenticated Codex access. It does not
require a GPU, a commercial EDA license, or a proprietary PDK. The evaluated
release is archived on Zenodo under DOI
\href{https://doi.org/10.5281/zenodo.21629308}{\nolinkurl{10.5281/zenodo.21629308}}, and the public repository is available at
\url{https://github.com/yuan-fd/DPLEvolve-AE}.

\subsection{Artifact Check-list (Meta-information)}

\begin{itemize}[leftmargin=*,itemsep=0pt,parsep=0pt,topsep=1pt]
    \item \textbf{Algorithm/program:} ReviewDSE performs white-box design-space exploration of OpenROAD detailed placement through a protected Teacher--Student workflow. Students make bounded changes to the source code, and the Teacher reviews full-flow results on nine targets and nine cut-row patterns.
    \item \textbf{Tools/build/binary:} The artifact builds pinned OpenROAD, ORFS, Yosys, and OpenSTA versions with GCC/G++~9 or later and CMake~3.16 or later. Each candidate is built as a separate OpenROAD binary.
    \item \textbf{Models/access/prompts:} The paper configuration uses one \texttt{gpt-5.5} Teacher and four \texttt{gpt-5.4} Students. The AE configuration uses one \texttt{gpt-5.6-sol} Teacher and one \texttt{gpt-5.6-terra} Student. All models use \texttt{xhigh} reasoning effort and are accessed through an authenticated Codex CLI session. The artifact includes all prompts. Other decoding parameters use the service defaults.
    \item \textbf{Data:} ORFS regenerates the Table~4 inputs. The artifact includes three checksummed source snapshots for Table~5 and a separate ${\sim}200$-MB archive of DEF, Verilog, and SDC files for Table~6.
    \item \textbf{Environment/hardware:} The workflow requires Linux x86-64, Bash~4 or later, Make~4 or later, Python~3.11 or later, and the standard OpenROAD build dependencies. The reference platform runs Rocky Linux~8.10 without a GPU and has two Xeon Platinum~8462Y+ processors (64 physical and 128 logical cores) and 314~GiB of RAM.
    \item \textbf{Execution/resources:} At least 100~GiB of free disk space is recommended. Setup takes several hours, and complete BO and model-search campaigns take several days; smaller machines may use less parallelism.
    \item \textbf{Metrics/output:} Runs report stage HPWL, strict legality, displacement, runtime, liveness, and source/binary/metric consistency. Scripts write fresh results, logs, review records, and figures.
    \item \textbf{Experiments/results:} The workflows reproduce Tables~4--6, Figures~4--5, selected-source replays, Level~1 calibration, and Level~2 search. Section~\ref{sec:ae-expected} gives the acceptance criteria.
    \item \textbf{Cost:} The paper reports an average of 2.15B logged tokens (${\sim}0.10$B active tokens) per target. Cost depends on the account and current provider prices.
    \item \textbf{Availability/licenses:} The public code uses the BSD 3-Clause License and contains no proprietary component. Third-party tools, data, and hosted models retain their own terms.
    \item \textbf{Workflow/archive:} GNU Make, Bash, Python, ORFS, and the Codex CLI manage the workflow. The evaluated archive is available under DOI \href{https://doi.org/10.5281/zenodo.21629308}{\nolinkurl{10.5281/zenodo.21629308}}.
\end{itemize}

\subsection{Description}

\noindent
\textbf{Access and dependencies.} The Zenodo archive and live repository provide
the same entry points and integrity checks. The OpenROAD baseline follows the
upstream \texttt{master} branch. The README gives the setup procedure, tested
tool versions, and exact pinned revisions. Setup requires Linux x86-64, Git,
rsync, the listed build packages, and network access, but neither root access
nor commercial software.

\noindent
\textbf{Data and models.} For Table~5 only, core utilization is set to 70, 90,
and 60 for AES, JPEG, and SWERV, respectively. Table~5 uses three archived
implementations across six selected and reference roles: AES uses LEGALM/Diamond, JPEG
uses Negotiation/Negotiation, and SWERV uses Diamond/Negotiation. Diamond and
Negotiation come from OpenROAD, while the agent implements LEGALM based on the
published method. Table~6 uses the paper's exact DEF, Verilog, and SDC
inputs.

The paper reports a search conducted in April--May~2026. Because Codex does not expose a stable
service revision, the harness records the model and reasoning effort; all
other decoding settings use their service defaults. The harness retries a
failed request up to eight times with a 45-s linear backoff. The repository
includes the prompts and corresponding request records. Because model outputs
are stochastic, repeated searches may produce different edits.

\subsection{Algorithm Flows}

Algorithm~\ref{alg:ae-level1} builds and freezes calibration evidence and
source starts; Algorithm~\ref{alg:ae-level2} uses that packet in a
target-specific Teacher--Student loop. Both call the protected evaluator, and
Level~2 cannot modify the packet.

\begin{algorithm}[H]
\caption{Level~1 calibration and warm-start construction.}
\label{alg:ae-level1}
\footnotesize
\begin{algorithmic}[1]
\Require Calibration cases $\mathcal{C}$, source $O$, evaluator $E$,
source starts $\mathcal{B}_0$, $S_C$ Students
\Ensure Frozen packet $W=(K,\mathcal{B})$
\State $K\gets\emptyset$; $\mathcal{B}\gets\mathcal{B}_0$
\ForAll{$c\in\mathcal{C}$}
    \State $b_c\gets\Call{Evaluate}{E,O,c}$; $Q_c\gets\Call{TeacherPlan}{c,b_c,K,\mathcal{B}}$
    \ForAll{$s\in\{1,\ldots,S_C\}$ \textbf{in parallel}}
        \State $O_s\gets\Call{Instantiate}{\mathcal{B},Q_c[s]}$
        \State $\Delta_s\gets\Call{StudentEdit}{O_s,Q_c[s]}$
        \State $\widehat{O}_s\gets\Call{Build}{O_s+\Delta_s}$
        \State $a_s\gets\Call{Evaluate}{E,\widehat{O}_s,c}$
    \EndFor
    \State $(K,\mathcal{B})\gets\Call{TeacherReview}{\{a_s\},K,\mathcal{B}}$
\EndFor
\State \Return $W\gets\Call{Freeze}{K,\mathcal{B}}$
\end{algorithmic}
\end{algorithm}

Each Student returns a complete source tree and its diff, together with the
full-flow metrics and corresponding evidence for legality, runtime, and liveness. The
Teacher reviews these results and freezes the selected evidence and source
starts for target search. Calibration uses AES, JPEG, and
SWERV at 70, 90, and 60 utilization, respectively.

During target search, Students work in private source trees while $W$ remains
read-only. A candidate is promoted only if it passes the build, legality,
metric, source-to-binary consistency, liveness, and runtime checks. The Teacher
retains other useful outcomes in target-local history $L_d$ without changing
$W$.

\begin{algorithm}[H]
\caption{Level~2 Teacher--Student target search.}
\label{alg:ae-level2}
\footnotesize
\begin{algorithmic}[1]
\Require Target $d$, source $O$, evaluator $E$, frozen packet $W$,
$S$ Students, $R$ iterations, runtime gate $g$
\Ensure Promoted candidate $p_d$ and target-local record $L_d$
\State $b_d\gets\Call{Evaluate}{E,O,d}$; $L_d\gets\{b_d\}$; $p_d\gets O$
\State \Call{CreatePrivateWorkspaces}{$S,O$}
\For{$r=1$ to $R$}
    \State $Q_r\gets\Call{TeacherPlan}{d,b_d,W,L_d,p_d}$
    \ForAll{$s\in\{1,\ldots,S\}$ \textbf{in parallel}}
        \State $O_s\gets\Call{Instantiate}{p_d,W,Q_r[s]}$
        \State $\Delta_s\gets\Call{StudentEdit}{O_s,Q_r[s]}$
        \State $\widehat{O}_s\gets\Call{Build}{O_s+\Delta_s}$
        \State $a_s\gets\Call{Evaluate}{E,\widehat{O}_s,d}$
    \EndFor
    \State $V_r\gets\Call{ContractValid}{\{a_s\},g}$
    \State $L_d\gets L_d\cup\Call{TeacherReview}{\{a_s\},V_r}$
    \State $p_d\gets\Call{PromoteBest}{V_r,p_d}$; \Call{PreparePeerEvidence}{$L_d$}
\EndFor
\State \Return $(p_d,L_d)$ \Comment{$W$ remains frozen}
\end{algorithmic}
\end{algorithm}

\balance
\subsection{Installation and Workflow}

From the artifact root, prepare the pinned environment and verify the toolchain
and model access as follows:
\scriptsize
\begin{verbatim}
make doctor
make bootstrap
make build-tools THREADS=8
make check
make prepare-paper-inputs CASE=aes_nangate45 THREADS=8
make validate-evaluator CASE=aes_nangate45 THREADS=8
make check-demo-models
\end{verbatim}
\normalsize
\texttt{make check-demo-models} sends one small request to each model and must
report \texttt{MODEL\_READY} for both roles. This verifies model access; the
functional evaluation does not require a live closed-loop search. Reproduce
the reported experiments as follows:
\scriptsize
\begin{verbatim}
bash artifacts/01-table4-qor/reproduce.sh \
  --threads 10
make check-table5-data
make reproduce-table5 THREADS=10
bash artifacts/03-table6-cutrow/reproduce.sh \
  --fetch --threads 10
bash artifacts/04-figures/reproduce.sh
make reproduce-paper-search \
  ACKNOWLEDGE_LLM_COST=yes THREADS=10
\end{verbatim}
\normalsize
The reproduction workflow runs the default flow on nine targets, performs 400
BO trials per target, replays 18 selected programs, evaluates the six Table~5
roles and 27 cut-row cases, and regenerates the figures. The workflow still
requires the configured Teacher and Student models. The final command starts a
new stochastic search and requires explicit acknowledgement of its model cost;
this paper-scale run is not part of the functional check.

\subsection{Evaluation and Expected Results}
\label{sec:ae-expected}

The scripts derive observed metrics from fresh runs before comparing them with
the references. No expected value is copied into an observed field.
\begin{itemize}[leftmargin=*,itemsep=0pt,parsep=0pt,topsep=0pt]
    \item \textbf{Table~4:} A full rerun produces nine legal cases. Mean HPWL changes are $-0.38\%$ for BO, $-1.78\%$ for HPWL selection, and $-1.68\%$ for GHR selection; ReviewDSE runtime ratios are $1.34\times$ and $1.11\times$. The acceptance tolerances are 0.06 percentage points for HPWL and 0.20 for each runtime ratio.
    \item \textbf{Table~5:} The six rebuilt role configurations form three selected/reference pairs, each of which must satisfy $\Delta H_{lg}<0$ and $\Delta H_f>0$. AES, JPEG, and SWERV report $-0.76\%/+1.48\%$, $-14.96\%/+20.96\%$, and $-0.12\%/+0.02\%$, respectively.
    \item \textbf{Table~6:} A full rerun produces 27 rows whose status and strict-legality classifications must match exactly; runtime may vary by 35\%. All nine ReviewDSE cases, one Diamond case, and one Negotiation case must be legal.
    \item \textbf{Figures~4--5:} Figure~4 must contain 96 observed points and mark the three unavailable SWERV points without imputation. Figure~5 must recompute the runtime-ratio Pareto set; the workflow writes fresh TSV and SVG files.
\end{itemize}
The public Level~1 profile reconstructs calibration because the exact Student
count and frozen packet are not available from the original run. A fresh search
is evaluated using its protected results and search trajectory; it is not
expected to reproduce the original source edits.

\subsection{Integrity and Failure Reporting}

Run the following checks before reporting any result:
\scriptsize
\begin{verbatim}
make test
make validate-configs
make zenodo-audit
\end{verbatim}
\normalsize
The audit checks all three Table~5 source snapshots. The Table~6 workflow starts
only after the downloaded archive passes its integrity check. Scripts write new results and
use references only for comparison. If a build, legality check, metric-contract
validation, or numerical comparison fails, the script exits with a nonzero
status and preserves the command, run record, result directory, and first
failure for inspection.

\subsection{Experiment Customization}

The \texttt{CASE}, \texttt{THREADS}, \texttt{STUDENTS}, and
\texttt{ITERATIONS} variables specify the target, level of parallelism, and
search size. Separate variables specify the Teacher and Student models. A new
campaign requires a distinct \texttt{DSE\_RUN\_PREFIX}. Candidate selection
always applies the paper's $2\times$ runtime gate, and paper-scale model runs
require the cost acknowledgement above.

\subsection{Methodology}

The artifact follows the current
\href{https://www.acm.org/publications/policies/artifact-review-and-badging-current}{ACM badging policy},
the \href{http://cTuning.org/ae/submission-20201122.html}{cTuning submission guide},
and the \href{https://github.com/ml-eda/artifact-evaluation/}{MLCAD AE instructions}.
\endgroup

\end{document}

%% file: setting-acm.tex
\iftrue
\fi

\usepackage{lipsum}
\usepackage{graphicx}
\usepackage{amsmath}
\usepackage{footnote}

\usepackage{amssymb}
\usepackage{mathtools}
\usepackage{mathrsfs}     
\usepackage{comment}
\usepackage[subrefformat=parens,labelformat=parens]{subfig}
\usepackage{bm,bbm}
\usepackage{multirow}
\usepackage{makecell}
\usepackage{threeparttable,booktabs}
\usepackage{blkarray}
\usepackage{tikz}
\usetikzlibrary{positioning,calc,fit,decorations.pathmorphing,shapes.geometric,shapes.gates.logic.US,calc}
\usepackage{balance}
\usepackage{courier}                                       
\usepackage{cleveref}                                      
\usepackage[mathcal]{eucal}
\usepackage[]{algpseudocode}                               
\algrenewcommand\textproc{\texttt}
\makeatletter\let\float@addtolists\relax\makeatother
\usepackage{algorithm}
\usepackage{filecontents}                                  
\usepackage{pgfplots}
\usepackage{pgfplotstable}
\usepackage{pgf-pie}
\usepgfplotslibrary{groupplots}
\pgfplotsset{compat=newest}
\usepackage[figuresright]{rotating}
\usepackage{xcolor,colortbl}                               

\theoremstyle{plain}

\theoremstyle{definition}

\algrenewcommand\textproc{\texttt}

\usepackage[skip=1pt]{caption}            
\definecolor{MYred}{RGB}{255,207,206}
\definecolor{MYsnow}{RGB}{213,248,241}
\definecolor{MYgreen}{RGB}{212,239,194}
\definecolor{MYpurple}{RGB}{199,195,239}
\definecolor{MYyellow}{RGB}{255,254,203}
\definecolor{MYorange}{RGB}{240,159,58}
\definecolor{MYblue}{RGB}{164,205,254}
\definecolor{CUHKorange}{RGB}{244,106,18} 
\definecolor{CUHKblue}{RGB}{0,111,190}    
\definecolor{CUHKgreen}{RGB}{0,127,128}   
\definecolor{CUHKred}{RGB}{228,46,36}     
\definecolor{CUHKyellow}{RGB}{198,148,34} 
\definecolor{CUHKdark}{RGB}{114,44,114}   
\definecolor{CUHKmiddle}{RGB}{144,44,144} 

\usepackage{tcolorbox}
\tcbuselibrary{skins,breakable}
    {\endtcolorbox}